\documentstyle[prb,aps,preprint]{revtex}

\begin{document}
\draft
\title{Kinetics of Electric field induced oxygen ion migration in 
epitaxial metallic oxide films}
\author{Arindam Ghosh\footnote{Electronic mail: aghosh@physics.iisc.ernet.in}}
\address{Department of Physics, Indian Institute of Science, 
Bangalore 560 012, India}
\author{A. K. Raychaudhuri\footnote{Electronic mail: arup@csnpl.ren.nic.in}}
\address{National Physical Laboratory, K.S.Krishnan Road,New Delhi 110012, 
India}
\author{R. Sreekala, M.Rajeswari and T. Venkatesan}
\address{Center for Superconductivity Research, University of Maryland, 
College Park, MD 20742}
                  
\date{\today}

\maketitle
\begin{abstract}

In this paper we report the observation of curent induced change of 
resistance of thin metallic oxide films. The resistance changes at a very 
low current (current density $J \geq 10^{3}$ A/cm$^{2}$). We find that the 
time dependence associated with the processes (increase of resistance) 
show a streched exponential type dependence at lower temperature, which 
crosses over to a creep type behavior at $T\geq$ 350 K. The time scale 
associated shows a drastic drop in the magnitude at $T \approx$ 350 K, 
where a long range diffusion sets in increasing the conductivity noise. 
The phenomena is like a "glass-transition" in the random lattice of 
oxygen ions.

\end{abstract}

\newpage

In recent years metallic oxides like LaNiO$_3$, RuSrO$_3$, 
La$_{0.5}$Ca$_{0.5}$SrO$_3$ etc. are being used as interconnects 
or electrodes in the field of oxide electronics. Many of these oxides 
belong to the ABO$_3$ class of oxides which are structurally similar
to high $T_c$ superconductors. In many of these oxides oxygen/oxygen 
defect is a particularly mobile species and the activation energy for 
diffusion is often $\leq$ 1eV.~\cite{ARIN1,ISLAM} The high mobility 
of oxygen species can lead to two generic problems in use of these 
oxides as interconnects or electrodes in electronic applications. 
First, the  conductivity noise in these oxides can be substantially 
large which mainly arises from long range diffusion of oxygen~\cite{ARIN1}. 
The second problem is that the current or field induced changes in
the resistivities of the films can occur for a prolonged period. This 
phenomenon of current induced resistance change have been first seen in 
high $T_c$ oxides and studied in films of YBa$_2$Cu$_3$O$_7$ in which
oxygen is a mobile species~\cite{MOECKLY}. It has been concluded that the 
current induced resistivity change arises due to  electromigration of 
oxygen ions. These problems are of concern because they can seriously 
limit their applications and cause reliability linked problems.

In this paper we investigated the kinetics of the electric field 
induced resistivity changes as a function of temperature in epitaxial thin 
films of a typical normal metallic oxide LaNiO$_{3-\delta}$ which has a 
cubic ABO$_{3}$ structure. The material is metallic in stoichiometric form  
($\delta$=0) and its electrical properties have been investigated
extensively~\cite{GAYA}. Our experiment consisted of high precision 
resistivity measurement as well as estimation of low frequency resistance
fluctuations. The investigation has led to three very important results  
which are of significance in determining the stability of the these oxides 
as interconnects or electrodes. The observations are : (1) a very small 
current and field can induce a long time drift in the resistivity of the 
film, (2) the time scales associated with the changes are strongly 
temperature dependent and (3) at $T \approx$ 340 K - 350 K there is a drastic 
reduction in the time scales associated with the process and the kinetics
of the resistivity variation changes over from a stretched exponential 
behavior (at lower temperatures) to a creep type behavior at higher 
temperatures. At these temperatures the conductivity noise also shows a 
rapid increase. 

All the experiments reported here are done with thin epitaxial films 
(thickness $\approx$ 150 nm) of LaNiO$_3$ grown on LaAlO$_3$ substrate  by 
pulsed laser ablation. Details of growth and characterization has been given 
elsewhere.~\cite{SATYA,SAGOI} Films with  room temperature resistivity    
$\rho_{rt}$ = 1.6 m$\Omega$.cm were patterned for resistance and noise 
measurements. The resistance  measurements were done with a precision of   
1 ppm and the noise measurement using a five probe geometry~\cite{SCOF} 
was done with precision of spectral power $\leq 10^{-19}$ V$^{2}$/Hz.
All the data are taken at thermal equilibrium where the temperature was 
controlled to within $\pm$5 mK.

In figure 1 we show a typical current  induced change in the observed 
resistance $R$ at three characteristics temperatures. The measurements shown 
here were made with a current of 100 mA. This corresponds to a current 
density $J \approx$ 5$\times$10$^4$ A/cm$^2$ and an electric field $E 
\approx$ 10$^2$ V/cm. $J$ was so chosen that it was not high enough to 
cause a rupture and at the same time not so low that no perceptible change 
occurs. (We find that there is a threshold associated with this phenomena 
with $J_{threshold}\leq$ 10$^{3}$ A/cm$^2$ and the exact value depends on 
the history of the material). Immediately after applying the current the 
resistance ($R$) drops slightly in a scale of few minutes. Then it starts to 
increase. We call this the  "damage"-($d$) process. Over several hundred minutes 
$R$ changes by about $\sim$ 2-3\%. Then the current was reversed. On reversal 
of current $R$ falls typically by $\leq$ 1\% and then starts to increase 
again. We call this drop in resistance the "recovery" ($r$) process. The 
time scale involved in the $r-$process is much smaller than that  involved 
in the $d-$process and it is quite similar to the early resistance drop 
seen on initial application of the current. The current stressing 
experiments were done for 270 K $<T<$ 400 K. Similar $r-$ and $d-$processes 
were seen in epitaxial films of high $T_{c}$ cuprates at room 
temperature~\cite{MOECKLY}. It was concluded~\cite{MOECKLY} that the 
resistance change is a consequence of the oxygen ion 
migration and while the $r-$process heals defects, the $d-$process increases 
the defect and disorder. In this letter however we donot discuss this issue 
and focus on the kinetics instead, in particular its temperature dependence.

It is also important to note that the process occurs at low current/field. 
The current density $J$ used in the experiment is much less than the typical   
electromigration threshold observed in metallic interconnects. Interestingly 
this is also less than the threshold of $J \approx$ 10$^{6}$ A/cm$^{2}$  
observed in high $T_{c}$ cuprates~\cite{MOECKLY}.
 
In order to obtain quantitative evaluation of the time scales we 
first fitted the data to the following  functional form. 

\begin{equation}
\Delta R(t) = \Delta R_{r}(t) + \Delta R_{d}(t)
\end{equation}

\noindent where the subscripts refer to the $r-$process and the $d-$process.
Here we assume that the two processes take place simultaneously. However  
they are well separated in time and thus it is possible to analyze the      
data unambiguously. Both $\Delta R_{r}$ and $\Delta R_{d}$ follow $streched$ 
exponential dependence on time $t$. For $\Delta R_{d}$ an additional creep 
component shows up at $T$ = 350 K and dominates the $t$ dependence at higher
$T$. Thus we can write :

\begin{equation}
\Delta R_{r} = \Delta R_{0r}
\left[1 - e^{-\left(t/\tau_r\right)^{\beta_r}}\right] 
\end{equation}

\noindent at all $T$.

\begin{equation}
\Delta R_{d} = \Delta R_{0d}
\left[1 - e^{-\left(t/\tau_d\right)^{\beta_d}}\right] + \nu t 
\end{equation}

\noindent where $\nu$ = 0 for $T <$ 350 K. At low  temperatures ($T <$ 330 K) 
$\tau_d$ being very large compared to the measurement time $t$ an
unambiguous and independent determination of $\tau_d$ and $\Delta R_{0d}$
is not possible. Instead, $\tau_d\gg t$ allows us to make the assumption
that 

\begin{equation}
\Delta R_{d} =  \left(\frac{t}{K}\right)^{\beta_{d}}
\end{equation}

\noindent where,

\begin{equation}
K = \frac{\tau_d}{\left(\Delta R_{0d}\right)^{1/\beta_{d}}}
\end{equation}

\noindent The data are fitted to the eqns.1-5 and the parameters are 
obtained. Typical time dependence of the resistance is shown figure 1. 
The inset to figure 1 shows a few fits to the data according to the 
equations above.

The time scale for the recovery process $\tau_{r}$ follows an Arrhenius 
relation with $T$ with an activation energy $E_{r}$ = 0.79 eV which is the 
same as the activation energy for migration of oxygen ions in these            
materials.~\cite{ARIN1,ISLAM} This shows that the process of current 
induced change in $R$ has its origin in oxygen ion migration. Also the 
exponent $\beta_{r} \approx$ 0.8 implying that the underlying process is 
not too different from a simple Debye relaxation ($\beta$ = 1). 

The temperature dependence of the time scale for the d-process, $\tau_{d}$ 
is shown in figure 2. In this case we show the temperature dependence of 
$K$ (see eqn 5) which in turn shows the T dependence of $\tau_{d}$ 
since the T dependence of $\Delta R_{0d}$ is less severe. We find that 
$K$ (and hence $\tau_{d}$) has a weak temperature dependence for $T <$ 340 K. 
But at $T$ = 350 K, $K$ drops drastically. In the same graph we have also 
shown the $T$ dependence of the creep rate $\nu$. This is small near 
$T$ = 350 K and increases rapidly at higher $T$. The creep rate also follows an 
Arrhenius temperature dependence with activation energy, $E_{d}$ = 0.84 eV. 
This again is similar to $E_{r}$ and is similar to the activation energy 
for oxygen ion migration. The physical process underlying the resistance 
change is thus related to the migration of oxygen ions. 

Figure 2 clearly shows that there is onset of long range diffusion at 
$T\geq$ 350 K which gives rise to the creep. The effect of long range   
diffusion for $T\geq$ 350 K can be seen also in the conductivity noise 
(denoted by $\gamma/n$, normalized noise magnitude, $n$ being the carrier
concentration) which we show in the lower part of figure 2. One can 
see that at the onset of the long 
range diffusion the temperature dependence of the noise shows an upturn at 
$T \approx$ 350 K. The fact that the  extra noise observed at $T\geq$ 350 K 
arises from long range diffusion has been shown before.~\cite{ARIN1}
Below this temperature long range diffusion is frozen-in. The situation 
is thus similar to the process of glass-transition where at certain 
characteristic temperature range the long range diffusion freezes on cooling.

Evidence of a glass-transition like freezing at $T \approx$ 350 K can be 
seen also in the $T$ dependence of the streched exponential exponent 
$\beta_{d}$ which is shown in figure 3. The value of the exponent 
$\beta_{d}$ is high in the "molten" state ($T >$ 350 K) signifying nearly 
single Debye type relaxation associated with the oxygen migration. 
However, in the frozen state ($T <$ 350 K) the value of $\beta_{d}$ sharply 
decreases to $\approx$ 0.4-0.6 implying that there is an hierarchy of the 
relaxation process as in a glass~\cite{BETA}. Interestingly, in a recent 
noise studies at low temperatures ($T <$ 20 K) we found clear evidence of 
two-level systems of low energy in this crystalline material as one 
generally find in conventional glasses~\cite{ARIN2}. In that study we 
were able to conclude that these "glass like" low energy excitations occur 
in the randomly frozen lattice of oxygen ions. The observation of kinetic 
freezing at $T \approx$ 350 K thus very well connects to the observation of 
low energy excitation.
 
The observation of a glass like freezing of the oxygen migration is a new 
result. (However, there exists evidence from calorimetric studies of a 
glass-transition like phenomena of the excess oxygen in a related 
compound La$_{2}$NiO$_{4}$~\cite{ITOH}.) We feel that this new phenomena 
will be extremely important in determining the  usability of oxide films 
as interconnects or electrodes. The important lesson from our investigation 
is that the characteristic temperature, where the "glass-transition" of 
oxygen ions occur should be at a fairly high temperature if these oxide 
interconnects have to be used in application with low noise and stability  
against field-induced oxygen ion migration.

\newpage
\centerline{\large\bf Figure Captions}
\vspace{50pt}

\noindent Figure 1: Time dependence of resistance at various temperatures.
The data at 400 K depicts creep-type behavior and the change in resistance 
is irreversible. The inset shows the type of fit that can be obtained with
the phenomenological equations described in the text. 

\vspace{30pt}
\noindent Figure 2: Temperature dependence of K (see text) and normalized 
noise magnitude. Current density for noise measurement was $\leq$ 10$^3$
A/cm$^2$. The dotted line in the noise curve is guide to the eye showing 
the expected temperature dependence due to localized movements of 
atoms/vacancies.

\vspace{30pt}
\noindent Figure 3: Temperature dependence of the exponent $\beta_d$ in
the long range diffusion regime.

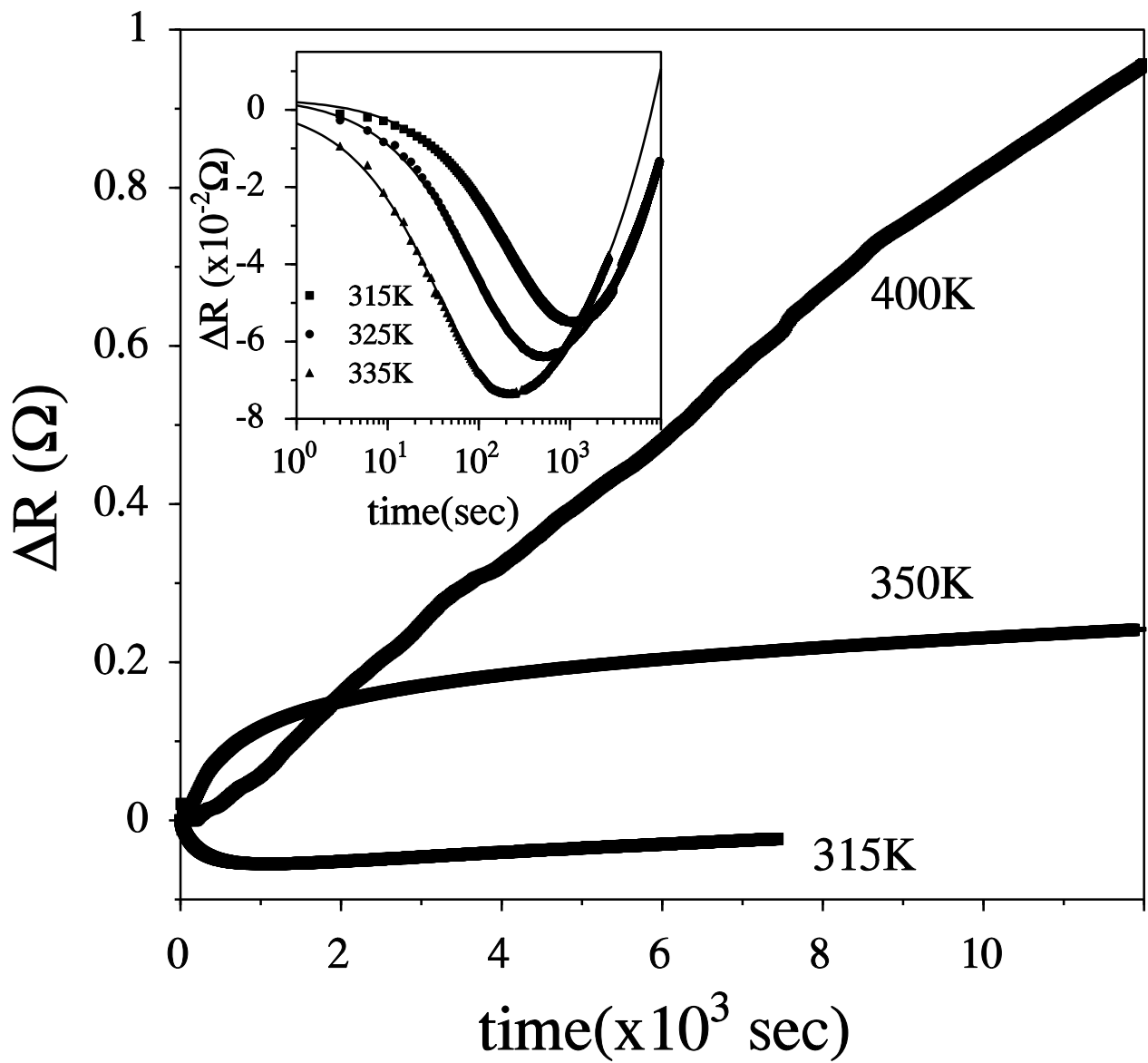
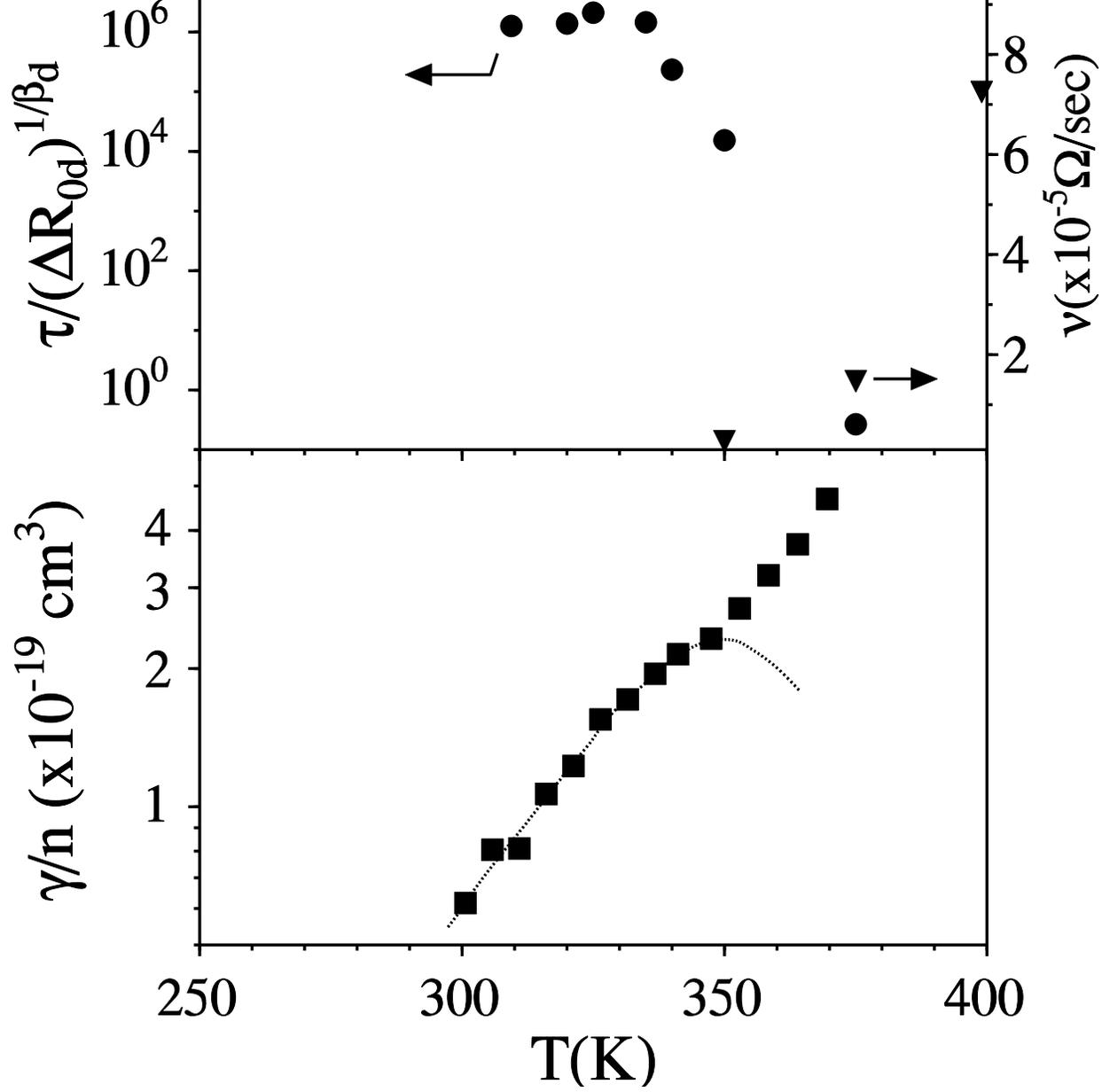
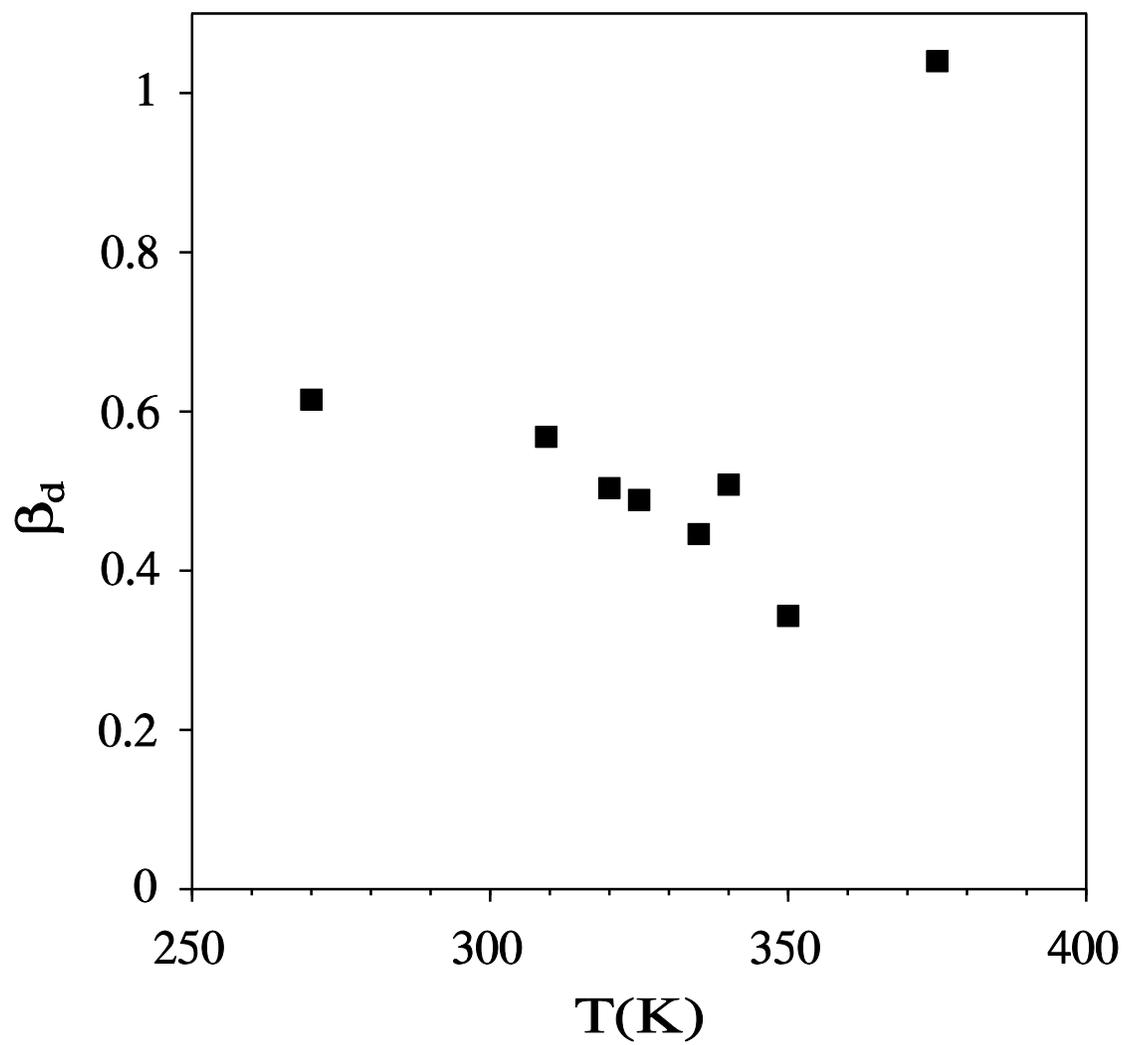
\end{document}